# MAKING CROSS-DOMAIN RECOMMENDATIONS BY ASSOCIATING DISJOINT USERS AND ITEMS THROUGH THE AFFECTIVE AWARE PSEUDO ASSOCIATION METHOD

John Kalung Leung[1], Igor Griva[2] and William G. Kennedy[3]

[1]Computational and Data Sciences Department, Computational Sciences and Informatics, College of Science, George Mason University, 4400 University Drive, Fairfax, Virginia 22030, USA

[2]Department of Mathematical Sciences, MS3F2, Exploratory Hall 4114, George Mason University,4400 University Drive, Fairfax, Virginia 22030, USA

[3]Center for Social Complexity, Computational and Data Sciences Department, College of Science, George Mason University, 4400 University Drive, Fairfax, Virginia 22030, USA

## ABSTRACT

*This paper utilizes an ingenious text-based affective aware pseudo association method (AAPAM) to link disjoint pseudo users and items across different information domains and leverage them to make cross-domain content-based and collaborative filtering recommendations. This paper demonstrates that the AAPAM method could seamlessly join different information domain datasets to act as one without any additional cross-domain information retrieval protocols. Besides making cross-domain recommendations, the benefit of joining datasets from different information domains through AAPAM is that it eradicates cold start issues while making serendipitous recommendations.*

## KEYWORDS

*Behavioral Analysis, Emotion-aware Recommender System, Emotion prediction, Personality, Pseudo Users Association.*

## 1. BACKGROUND

Researchers often encounter situations in working with multiple datasets in the same or different information domains. A disjoint object occurs when an object identifier (id) of a data file in a dataset makes a cross-reference to another dataset of an object having the same object id. Both objects are individual and different in that they are not the same object, except they happen to have the same id assignment. Both objects may reside in the same or different information domains.

This paper intrigues by [1], which has examined a similar situation about disjoint users of a data file when making cross-reference to different datasets within the domain. In [1], it developed a method known as an Affective Aware Pseudo Association Method (AAPAM) to Pseudo Association Connect (PAC) disjoint users reside in different datasets of the same domain.

Illustrate below is a disjoint user example. A user U assigned with id 123 in data file F of dataset S in domain D is not the same user U assigned with id 123 in data file G of dataset T in same domain D. However, user V assigned with id 987 in data file G of dataset T in the same domain D could be the same user U in data file F of dataset S in domain D.

In contrast, a pseudo disjoint user refers to a user says user U and user V. Both are individual users with a closely similarly emotion profile, so that for grouping users, they are grouped because both have a similar emotion profile. Both U and V users are considered pseudo disjoint users. AAPAM method provides a way to PAC connect disjoint users pairwise by the Affective Index Indicator (AII) values of disjoint users. AII value obtains by computing the Cosine Similarity of the pairwise disjoint users' emotion vector (UVEC). When AII is one (1), it represents the pair of disjoint users with identical emotion profile values. AII with unity value represents a high certainty the pair of disjoint users are the same user in a virtual sense, except with different user id assignment and reside in different datasets. Such PAC reconnection made for the pair disjoint users is known as a pseudo association. PAC connect disjoint users do not need to have a unity AII value. One can set a threshold value on AII; for example, 98%, any disjoint user's AII meets the threshold and becomes a member of the pseudo-user group, a group of disjoint users who share closely similar AII values as expressed in their UVEC.

## 2. MOTIVATION

Paper [1] developed the AAPAM method and explained what it is as a proof of concept to show users with the same user id that reside in different datasets within the same domain are disjointed and use the PAC method to pseudo associate disjoint users together. However, in [1], it did not mention how to use AAPAM and PAC in Recommender for making top-N recommendations. This paper wants to expand the AAPAM, and PAC works of [1] to generalize disjoint objects and determine whether the AAPAM method can PAC disjoint objects across different datasets and information domains. This paper will follow the procedure described in paper [1] to prepare emotion profiles for all involved datasets under study.

Once the AAPAM and PAC.methods have proved to work with disjoint users reside in different datasets across different information domains, this paper would like to show it is possible to build a Collaborative Filtering Cross Information Domains Recommender System (CF-CID RS) to make top-N recommendations using affective enabled datasets from different information domains. This author believes the CF-CID RS effort is worthwhile because there is little work in the user's mood-oriented RS field addressing a Recommender capable of adapting user's moods in making top-N recommendations over CID datasets.

## 3. INTRODUCTION

### 3.1. MovieLens Datasets

Paper [1] works with four MovieLens datasets [2], namely, ml-latest-small (a.k.a. mlsm), ml-20m (a.k.a. ml20m), ml-25m (a.k.a. ml25m), and ml-latest (a.k.a. ml27m hereafter) datasets. Table 1 illustrates the statistics of each mentioned MovieLens dataset. The name of a MovieLens dataset reflects the number of ratings it contains.

Table 1 Statistics of MovieLens datasets.

| Attribute Dataset | No. Users | No. Movies | No. Ratings | No. Overviews |
|---|---|---|---|---|
| mlsm | 610 | 9742 | 100836 | 9625 |
| ml20m | 138493 | 27278 | 20000263 | 26603 |
| ml25m | 162541 | 62423 | 25000095 | 60494 |
| ml27m | 283228 | 58098 | 27753444 | 56314 |

In each of the four MovieLens datasets, there are two data files named ratings, and tags contain user ids as a unique identifier. MovieLens stated that user ids found in ratings and tags data files are consistent within the same dataset but are not uniform across different datasets [3]. For example, in [1], user id 400 is only compatible within the same MovieLens mlsm dataset and is not across ml20m, ml25m, ml27m datasets. In other words, user id 400 in other MovieLens datasets are not the same user id 400 as in the mlsm dataset. However, [1] has demonstrated by using the AAPAM method, the disjoint user id 400 in mlsm can correctly connect to the proper user id in MovieLens datasets as depicted in Table 2.

## 3.2. Affective Aware Pseudo Association, Pseudo Association Connection, and Affective Index Indicator

The Affective Aware Pseudo Association Method (AAPAM) computes the Affective Index Indicator (AII) using the Cosine Similarity algorithm [4], as depicted in Equation 2, to express the closeness of the emotion profiles between two users or items. When using AAPAM to compare pairwise between User id 400 of mlsm against users in other MovieLens datasets, AII reveals, as depicted in Table 2, the closest other users' emotion profiles that match the candidate user. User id 400 in mlsm can make a pseudo associate connection (PAC) to user id 66274 with AII 0.999916 in ml20m, or to user id 95449 with AII 0.999999 in ml25m, or user id 89195 with AII 0.999999 in ml27m, respectively.

$$Inner(x, y) = \sum_i x_i y_i = <x, y> \qquad (1)$$

$$CosSim(x, y) = \frac{\sum_i x_i y_i}{\sqrt{\sum_i x_i^2} \sqrt{\sum_i y_i^2}} = \frac{<x, y>}{||x|| ||y||} \qquad (2)$$

AAPAM also worked with The Movie Database (TMDb) [5], where the movie metadata has been scraped from TMDb for movie overviews, poster images, and other metadata. AAPAM applied the Tweets Affective Classifier (TAC), a method developed in [6] to classify a movie emotion profile. A movie emotion profile is also known as a movie vector, mvec, which represents a multi-dimensional embedding of a probability distribution of seven primary human emotions: neutral, happiness, sadness, hate, anger, disgust, and surprise. Each user in [6] also has a user emotion profile, uvec, where it contains the average value of all movies mvecs the user has watched.

Table 2 Pseudo Association Connection of ml-latest-small user id 400 to other users in different datasets through affective index indicator.

| Dataset | mlsm | ml20m | ml25m | ml27m |
|---|---|---|---|---|
| User1 ID PAC | 400 | 66274 | 95459 | 89195 |
| User1 Movie Count | 43 | 22 | 43 | 43 |
| User1 Watched List movieID | 6<br> 47<br> 50<br> 260<br> ...,<br> 122886<br> 134130<br> 164179<br> 168252 | 47<br> 260<br> 300<br> 307<br> ...,<br> 2628<br> 2797<br> 3418<br> 3481 | 6<br> 47<br> 50<br> 260<br> ...,<br> 122886<br> 134130<br> 164179<br> 168252 | 6<br> 47<br> 50<br> 260<br> ...,<br> 122886<br> 134130<br> 164179<br> 168252 |
| User1 UVEC<br>Neutral<br>Happiness<br>Sadness<br>Hate<br>Anger<br>Disgust<br>Surprise | 0.16353<br>0.08874<br>0.12709<br>0.20332<br>0.11934<br>0.15881<br>0.13918 | 0.16250<br>0.08609<br>0.12654<br>0.20701<br>0.11776<br>0.16005<br>0.14005 | 0.16353<br>0.08874<br>0.12709<br>0.20332<br>0.11934<br>0.15881<br>0.13918 | 0.16353<br>0.08874<br>0.12709<br>0.20332<br>0.11934<br>0.15881<br>0.13918 |
| User1 Affective Index Indicator | 1.0 | 0.99992 | 0.99999 | 0.99999 |

As illustrated in Table 2, user id 400 in the rating data file of the mlsm dataset has watched 43 movies; taking the average of all the 43 movies' mvecs yields the uvec for user id 400. As mentioned in [6], a movie's mvec is static and stays unchanged throughout the film's life; whereas, a user's uvec changes its value each time the user watches a movie. The user's uvec reflects the up-to-date movie taste and preference of the user. A movie mvec is unique, while a uvec may not be unique when two users watched the same movie set.

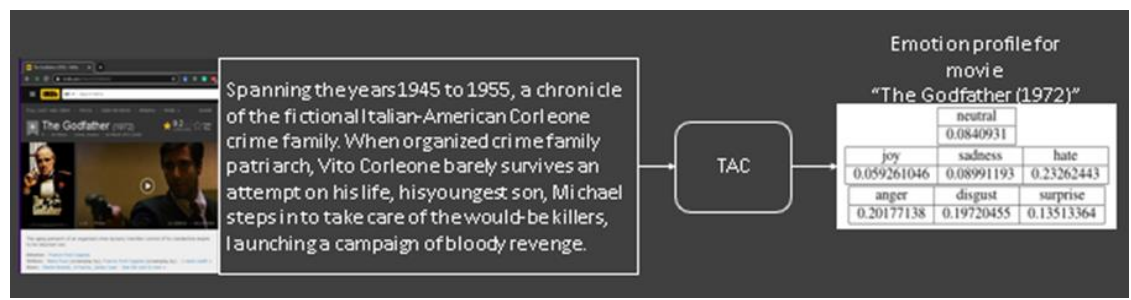

Figure 1. Emotion profile of film, "The Godfather (1972)", Classified by TAC.

Paper [6] advocates that every movie can obtain an emotion profile through TAC by classifying the movie's moods over the corresponding movie overview. Figure 1 shows an example of the movie, "The Godfather (1972)", which rated by the Internet Database (IMDb) portal as the top movie of all time [7], has its emotion profile, mvec, classified by TAC using the movie's overview as input. As shown in Figure 1, the movie's overview feeds as input to TAC yields the movie emotion profile, mvec. By ranking the probability distribution values of mvec in the descending order, it yields "hate," a.k.a. "fear" in this paper, as the dominant mood for the example movie, follows by "anger," "disgust," "surprise," "sadness," "neutral," and "joy" a.k.a. happiness in this paper. One can perform the affective analysis for the film by describing the movie in the following way. The film "The Godfather (1972)" is a movie with plots full of hate. The film is an angry movie. It is a disgusting film due to its violent content.

Nevertheless, the movie's plots are full of surprises, but overall a sad film. No viewer would say, "The Godfather (1972)" is a joyful film. By merely reciting the ranking of moods in mvec, one explains the movie. Similarly, one can look at a user's uvec who has watched the film feel how well the user liked it.

Besides, the AAPAM method can PAC connect disjoint users from different datasets within the same domain; this study believes the same technique can PAC connect disjoint users and items among different datasets across different domains. Unlike MovieLens datasets, some other movie datasets such as TMDb highlight the average voting score, the sentiment rating value on a scale of 1 (lowest) to 10 (highest), of a movie by a group of users who have watched and rated the movie through the voting count attribute in the data file instead of individual user's sentiment. TMDb does not maintain user information, and neither contains a user-id field in the dataset. When applying the AAPAM to connect MovieLens and TMDb domains, the PAC connection applies to movie items between MovieLens and TMDb. Here, the PAC connection between movie A in MovieLens to movie B in TMDb indicates how similar the two movies' emotion profiles, mvecs, form a one-to-one relationship. However, when applying the PAC to connect a user in MovieLens and a movie item in TMDb, the movie item mvec in TMDb must first be normalized with the respective voting count. The normalized mvec represents the average uvec of the group of users who have rated it. Thus, the PAC connection between user A in MovieLens to the normalized mvec of movie B in TMDb indicates how similar user A to a group of users B is in the form of a one-to-many relationship.

### 3.3. Amazon Review Data Datasets

In the recent release of the updated version Amazon Review Data (2018) repository [8], [9], and [10], the repository maintains product reviews and product metadata on 29 categories stored in two data groups: Complete-review-data and Small-subsets-for-experimentation. This study uses the following two data files in the Small-subsets-for-experiment group of 5-core: the reviews_Toys_and_Games_5.json, a.k.a Toys_and_Games hereafter, and the Digital_Music_5.json, a.k.a. Digital Music hereafter. The data size of Amazon Review Data (2018) contains 233.1 million reviews in the Complete-review-data group spanning across 29 product categories from May 1996 to October 2018. Some categories have fewer than 90,000 to 900,000 reviews, while many have over one million to more than 51 million reviews. Also available for experimentation are subsets of smaller datasets with reviews ranging from a few thousand to a few million, as depicted in Table 3 in JSON format [11].

Amazon Review Dataset contains reviewerID and reviewerName, which can be combined to form a unique user id. The ASIN code stands for Amazon Standard Identification Number, represents a unique product code to use as an item id. The vote field contains a reviewer rating score on a 1 (lowest) scale to 5 (highest). The reviewText contains subjective writing of the reviewer's sentiment, a source for emotion classification for the item emotion profile, ivec. In the case of Amazon Review Dataset, a user's emotion profile, uvec, represents the average value of all items' ivecs the user has reviewed. Note that ivec in Amazon Review Dataset is equivalent to mvec in MovieLens.

Table 3. A sample review extracted from Amazon Review Dataset.

```
{
  "reviewerID": "A2SUAM1J3GNN3B",
  "asin": "0000013714",
  "reviewerName": "J. McDonald",
  "vote": 5,
  "style": {
    "Format:": "Hardcover"
  },
  "reviewText": "I bought this for my husband who plays the piano.  He is having
a wonderful time playing these old hymns.  The music  is at times hard to read
because we think the book was  published for singing from more than playing
from.  Great purchase though!",
  "overall": 5.0,
  "summary": "Heavenly Highway Hymns",
  "unixReviewTime": 1252800000,
  "reviewTime": "09 13, 2009"
}
```

## 3.4. Scope of Work

This study intends to show it is possible to make Collaborative Filtering [12] and [13] recommendation of items to an active user once the AAPAM method has proved to work for PAC connecting disjoint users and items in MovieLens's mlsm dataset to Toys_and_Games, and Digital_Music data files in Amazon Review Dataset. In [6], [14], and [1], these papers had examined issues related to recommendations making by affective aware recommenders. The Recommenders involved in making recommendations were all within the same domain confined by the MovieLens datasets. This paper aims to demonstrate any Recommenders, besides making recommendations the usual way within the confined domain, also can make cross-domain recommendations by embracing the AAPAM technique. For example, a cross-domain recommendation may appear in the following situations: "People who enjoyed movie T in MovieLens may choose to play V in Toys_and_Games and listen to W in Digital Music." Another example of cross-domain recommendation between a user in MovieLens and a group of users in TMDb may appear as follow: "People who enjoyed movie X in MovieLens may also enjoy movie Y of the similar taste of group users Z in TMDb." Using the AAPAM method to make cross-domain recommendations for disjoint users and items in different datasets across different domains is the contribution this paper makes, and this investigator team may be the first to perform such a study.

## 4. RELATED WORK

This study aims to object emotion profile (EP) modeling for users and items and cross information domains (CID) Recommenders on top-N recommendations making. This section examines the prior works performed on EP modeling before surveying the CID Recommenders.

## 4.1. Emotion Profile Modeling

Recommender researchers recognize that human emotions are fundamental to daily human activity [15]. Emotions also play a critical role in modeling human preferences and influencing human decision-making processes. Nevertheless, emotion aware Recommender (EAR) is still a wide-open field for research [16]. Researchers use facial expressions developed by Paul Ekman in his cross-cultural human emotions research [17] as the preferred method to detect and recognize primary human emotion. Although Ekman's facial expressions offer a mature methodology to detect and recognize emotional features, it fails in the absence of a faceless object.

The lack of a standardized method in classifying types of emotions causes researchers to struggle to agree on the underlying emotional experiences and phenomena in studying human emotional experiences and expressions [18] and [19]. Some researchers approach understanding emotional phenomena by developing techniques and methodologies in analyzing Drosophila and other insects' behaviors. They found four types of raw emotions: happiness, sadness, anger, and fear, which are associated with three core effects reward (happiness), punishment (sadness), stress (anger and fear) [20], which coincide with the three primary colors red, blue, and yellow that when combining the colors in various proportions will yield more complex emotions, for example, aesthetic and love emotion. Gu et al. in [21] proposed to name the color emotion scheme as the "Three Primary Color Model of Basic Emotions." A recent human primary emotion research by the Institute of Neuroscience and Psychology at the University of Glasgow reported only four primary emotions: happiness, sadness, fear/surprise, and anger/disgust [22]. The number of primary human emotions matches what had been reported earlier [20]. Other researchers, such as Robert Plutchik, the inventor of the Wheel of Emotions, advocated eight primary human emotions that include happiness, sadness, trust, anger, fear, anticipation, disgust, and surprise [23].

## 4.2. Cross Information Domains Recommenders

For a Recommender to make a top-N recommendation, it must access the information for data processing. To make a CID top-N recommendation, a Recommender must integrate information from a target domain with information in the source domain and consistently process the information in both domains as if both domains are one. In [24], it defined two tasks a Recommender must be capable of to make CID top-N recommendations:

1) to improve the quality of making top-N recommendations in target domain B, one can exploit information about users and items in the source domain A.

2) the Recommender can make joint top-N recommendations for items that are members of different domains; that is, recommending joint items to joint users in both joint source and target domains.

However, [24] did not exploit any contextual information of an item and only work with item attributes metadata that is closely related, such as rating score in the source domain and weight value in the target domain. Moreover, [24] did not model user profile adaptively to reflect a user's tastes and moods change over time. Nevertheless, it proposed a taxonomy to classify CID recommendation methodologies. It classified Recommender filtering into two classes: Content-based filtering (CBF) and Collaborative filtering (CF) and identified the relationships between domains against the Recommender filtering classes. Relationships between domains for CBF included attributes, social tags, semantic properties, and correlations, while for is ratings, rating patterns, latent factors, and correlations. Moreover, [24] defined recommendation models into the adaptive and collective models.

[25] Advocated the role of emotions in the context-aware Recommender is to adapt users' preferences across different contexts. The paper pointed out that emotion is one of the most important contextual features but receives little attention from researchers. The paper evaluated two types of context-aware recommendation algorithms - context-aware splitting and differential context modeling as a new strategy to aggregate multiple users' tastes in generating group recommendations.

## 5. METHODOLOGY

This study applies affective features to three data sources that it uses: users' emotion profiles, uvecs, and items' emotion profiles, ivecs, or mvecs equivalent. No disjoint users and items can interconnect without adding affective features across data sources of MoieLensm TMDb, and Amazon Review Dataset. Affective aware features are added to the data sources through Tweets Affective Classifier (TAC) as developed in [6]. Table 2, Table 4, and Table 5 depict samples after added affective features uvec and mvec in this study's data sources.

Table 4. TMDb movie emotion profile example.

| tmdbId | 2 | 525662 |
|---|---|---|
| movieId | 4470 | 189111 |
| Mood | Disgust | Hate |
| Neutral | 0.15705037 | 0.11876434 |
| Happiness | 0.08608995 | 0.05086204 |
| Sadness | 0.15583897 | 0.12669845 |
| Hate | 0.07506061 | 0.3391073 |
| Anger | 0.08469571 | 0.13069303 |
| Disgust | 0.26612538 | 0.13746719 |
| Surprise | 0.17513901 | 0.096407644 |

Table 4. Amazon Review Dataset of Digital Music from Small Subsets for Experimentation Group.

| userId | 8129 | 13878 |
|---|---|---|
| Neutral | 0.1976067315 | 0.1826606686 |
| Happiness | 0.137053136 | 0.1766339161 |
| Sadness | 0.1214098371 | 0.1509535014 |
| Hate | 0.0877815959 | 0.0599035051 |
| Anger | 0.0799363965 | 0.0571205363 |
| Disgust | 0.1656119045 | 0.1604166086 |
| Surprise | 0.2106004065 | 0.2123112464 |

### 5.1. Tweets Affective Classifier

Tweets Affective Classifier (TAC), an emotion classifier developed in [6] tweet text. When applying TAC to TMDb's movie overview, it classifies the mood of the overview, which becomes the movie's emotion profile, mvec, as depicted in Table 4. After joining the movie emotion profile in the TMDb data file with the ratings data file in MovieLens, users' emotion profiles, uvecs can be computed by taking the average of mvecs from all the movies a user has watched. Similarly, TAC can be applied to users' reviews in any ratings data file of Amazon Review Dataset to obtain users' emotion profiles, uvecs, as depicted in Table 5.

TAC is built from an asymmetric butterfly wing double-decker bidirectional LSTM - CNN Conv1D architecture [6], [26], and [27] as depicted in Figure 2, Figure 3, and Figure 5 in block diagrams for the classifier to detect and recognize emotional features from tweets' text messages. A preprocessed seven emotional word embeddings were applied to train TAC through pre-trained GloVe embeddings using the glove.twitter.27B.200d.txt dataset [28]. There are two types of input word embeddings: trainable emotion words embeddings and frozen emotion words embeddings, i.e., the weights in the embeddings are frozen and do not allow for modification during TAC's training session.

The training process begins with training the first half of the butterfly wing by feeding preprocessed TAC input emotional word embeddings to the double-decker bidirectional LSTM neural nets. The frozen emotional word embeddings are fed to the top bidirectional LSTM, while the trainable emotional word embeddings feed to the bottom bidirectional LSTM. Next, the top and bottom bidirectional LSTM of the double-decker neural net are concatenated. The double-decker bidirectional LSTM is fed in parallel to seven sets of CNN Conv1D neural nets. The dropout regularization unit with a parameter set at 0.5 to prevent overfitting. All the Conv1Ds outputs are concatenated to form the first half of the butterfly wing neural nets' overall production.

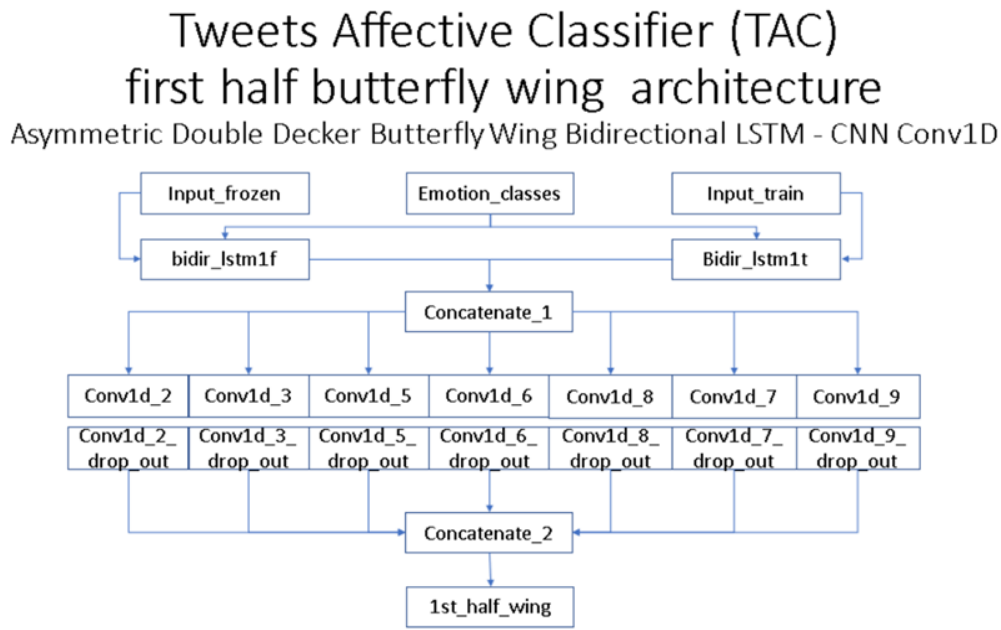

Figure 2. First half of Asymmetric Butterfly Wing Double-decker Bidirectional LSTM-CNN Conv1D architecture block diagram of Tweets Affective Classifier.

The second half of the butterfly wing neural nets architecture is different from the first. The training process of the second half wing begins by setting up two groups of Conv1D neural nets. Each group contains seven CNN Conv1D neural nets. The preprocessed TAC's frozen emotional word embeddings are fed as input in parallel to group one of the Conv1D neural net while feeding in parallel to the other group's trainable emotional word embeddings. The dropout regularization parameter is set at 0.5 for all seven pairs of conv1ds to prevent overfitting.

Concatenation of all seven pairs of Conv1D outputs becomes a single output and fed to a single bidirectional LSTM with the dropout regularization parameter set at 0.5 to prevent overfitting.

Next, by concatenating the first half of the butterfly wing's output with the second half, it produces the overall output. The output then feeds to a MaxPooling1D with the dropout regularization value set at 0.5. Then the flow goes through a Flatten neural net before flowing through a Dense neural net. Finally, the process flows through another Dense neural net using sigmoid as an activation function to classify the emotion probabilistic distribution values. TAC output the emotion classification in the form of the probabilistic distribution of seven values that sum to one (1). Each value indicates the amount in the percentage of the emotion class. Thus, the seven-emotion probabilistic distribution value forms an emotion profile of an object.

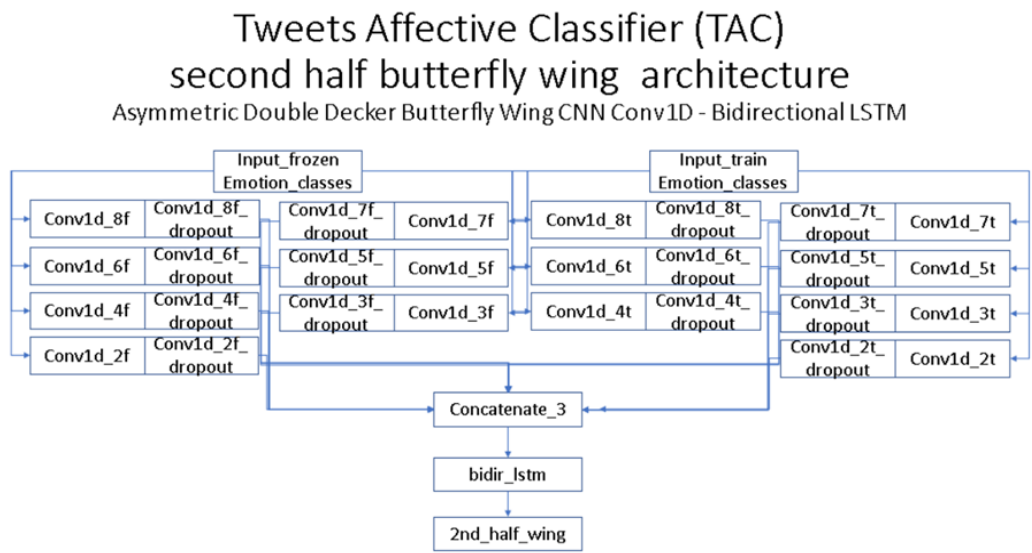

Figure 3. Second Half of Asymmetric Butterfly Wing Double-decker Bidirectional LSTM-CNN Conv1D Architecture Block Diagram of Tweets Affective Classifier.

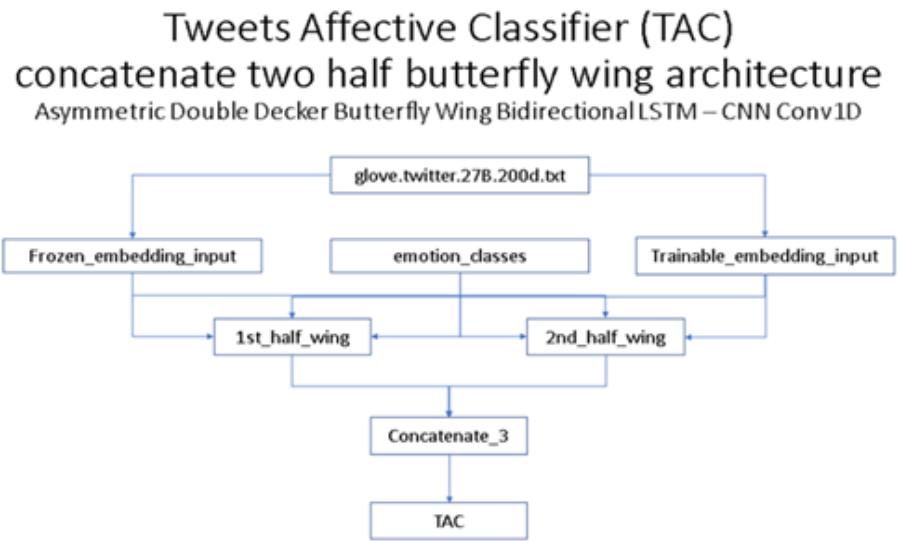

Figure 4. Concatenate two half butterfly wing architecture block diagram of Tweets Affective Classifier.

## 5.2. PAC Disjoint Users within Same and Across Different Domains

A word of caution is on user id that appears across different MovieLens datasets. MovieLens collects datasets anonymously. Thus, there is no guarantee that a user id that appears in one dataset is the same person who bears the same user id in a different dataset. Using user id to make cross-reference metadata across different MovieLens dataset may not produce a consistent result. In this study, the mlsm dataset designates as the training dataset. The other datasets: ml20m, ml25m, and ml27m, are concatenated into a large dataset uses for testing.

To overcome the disjoint user's deficiency across different MovieLens datasets, an affective aware pseudo association method (AAPAM) deploys to associate disjoint user across different datasets. The pseudo associate connection (PAC) connects disjoint users across various datasets method works as follow:

- For each user id in a MovieLens dataset, computes the user's uvec.

- Concatenate ml20m, ml25m, and ml27m datasets together as a massive dataset use for testing. Note that all user id in the testing dataset should have its uvec computed.

- For each user id in the training dataset, compute the pairwise Cosine Similarity for the training user id against each user id in the testing dataset.

- The Cosine Similarity result obtained from the pairwise user id will go through a ranking and sorting in descending order. The top pairwise user id on the sorted list, indicated by the AII value, has the closest match of emotion profile between the training user id and the testing user id. PAC connects the most similar pair of training user id and testing user id.

- Table 2 depicts user id 400 of mlsm PAC connects to other user ids across Movielens datasets.

A careful examination of Table 2 regarding the PAC of user id 400 in the mlsm dataset to the other three larger MovieLens datasets, user id 400 in mlsm PAC to user id 66274 in ml20, user id 95459 in ml25m, and user id 89185 in ml27m, the Cosine Similarity between user id 400 in mlsm or its AII value and other PAC users is virtually identical. Except for PAC users in ml20m, the other PAC users' AII values in ml25m and ml27m are identical. Both PAC users in ml25m and ml27m have the same history of movie watched list as user id 400 in mlsm. Thus, their uvecs are correctly aligned. From an observer perspective, all three users are the same in the MovieLens dataset.

The AAPAM scheme works. The method provides a means to associate disjoint users across different datasets within the same domain. However, PAC mlsm user id 400 with other disjoint users in ml20m, ml25m, and ml27m datasets does not offer any additional value-added information. The method failed to enlarge the extra data point for user id 400 in mlsm.

Instead of selecting a low movie watched count user as a test user, such as user id 400 in mlsm who only has watched 43 films, a high film viewing count user is favorable for the test user. In the mlsm dataset, the following users have movie watched count in the range of 1,300 to 2,700. Depicted in Table 7 are selected test users with respective PAC information. Table 7 contains seven columns. The first column represents the identity of user 1 to user 3. For example, row U1 shows user 1 with id 414 in the mlsm dataset can PAC to user id 125022 in the ml20m dataset. Alternatively, user id 414 can PAC from mlsm to user id 131662 in the ml25m dataset.

Moreover, the same user can PAC from mlsm to user id 236165 in the ml27m dataset. The above PAC actions are all involving disjoint users in the same MovieLens domain. However, columns 6 and 7 of Table 7 show user id 414 has a choice to PAC from mlsm dataset in MovieLens to user id 10354 of Digital Music dataset or user id 93437 of Toys and Games dataset in Amazon. Indeed, besides capable of connecting disjoint users from different datasets within the same domain, PAC can also connect disjoint users from different datasets across different domains. There are rows in Table 7 that contain rated movies counts for each user in its respective dataset. Also, rows holding movies watched list, uvecs, and the AII values indicate individual users' Cosine Similarity.

## 6. IMPLEMENTATION

### 6.1. Making Cross Domain Recommendations

The paper in [14] illustrated a strategy of making recommendations through a Comparative Platform implemented with various affective aware Recommenders within the MovieLens domain. Table 6 depicts the five top-20 recommendations list generated by the five respective Recommenders in the Comparative Platform. The five Recommender algorithms deployed in [14] are Item-based Collaborative Filtering Recommender, User-based Collaborative Filtering Recommender, Genres Aware Recommender, Emotion Aware Recommender, and Multi-channel Affective Recommender. Moreover, in the article [1], a follow-up paper of [14], illustrated the validation method's development for recommendations made by the Comparative Platform Recommenders. As an example, Table 6 depicts the top-20 recommendations for the user id 414 made by the five Recommenders in the Comparative Platform Recommenders.

Table 6. Top-20 Recommendations List Generated by Recommenders in Comparative Platform for Test User ID 414.

| U414 | IBCF | UBCF | GAR | EAR | MAR |
|------|------|------|------|------|------|
| 1 | 1291 | 1258 | 73499 | 858 | 131714 |
| 2 | 1196 | 8368 | 3030 | 53519 | 129229 |
| 3 | 260 | 2424 | 4565 | 363 | 704 |
| 4 | 1270 | 1230 | 3389 | 3109 | 112911 |
| 5 | 1210 | 1982 | 1049 | 3330 | 1606 |
| 6 | 2115 | 3176 | 2153 | 6981 | 3389 |
| 7 | 2571 | 1219 | 809 | 3211 | 2421 |
| 8 | 1240 | 48385 | 7925 | 5569 | 3704 |
| 9 | 1197 | 34542 | 131714 | 5853 | 442 |
| 10 | 1036 | 1449 | 112897 | 4224 | 4367 |
| 11 | 1136 | 6879 | 27837 | 4608 | 87430 |
| 12 | 1200 | 42418 | 1867 | 135133 | 2826 |
| 13 | 1214 | 520 | 1606 | 109673 | 1049 |
| 14 | 2716 | 1183 | 704 | 2009 | 122922 |
| 15 | 4993 | 2788 | 56775 | 110655 | 64695 |
| 16 | 1265 | 1693 | 91485 | 29 | 2683 |
| 17 | 2028 | 40732 | 115727 | 4433 | 3702 |
| 18 | 7153 | 2324 | 32511 | 2256 | 131739 |
| 19 | 3578 | 1333 | 6664 | 3028 | 76743 |
| 20 | 858 | 2728 | 761 | 2450 | 57326 |
| Hit % | 90% | 65% | 20% | 35% | 30% |

AAPAM demonstrates its ability to PAC connect disjoint users across the same or to a different domain. By integrating Recommender with the PAC capability, the Recommender can make recommendations across a different domain; for example, PAC connects disjoint users from the MovieLens domain to users in the Amazon domain The PAC connects disjoint users because they share similar affective features, and their respective uvec closely match each other. Table 8 shows the result of a Recommender making a cross-domain recommendation between the MovieLens and Amazon domains.

## 7. EVALUATION

Table 7 depicts the PAC connections of disjoint users in MovieLens to Amazon Digital Music (MUS) and Toys and Games (T&G) datasets.

Table 7. PAC Connect Disjoint Users from MovieLens to Users in Amazon.

| Dataset | mlsm | ml20m | ml25m | ml27m | MUS | T&G |
|---|---|---|---|---|---|---|
| U1 | 414 | 125022 | 131662 | 236165 | 10354 | 97437 |
| U2 | 448 | 63555 | 134534 | 182133 | 15262 | 206109 |
| U3 | 474 | 96370 | 107581 | 54271 | 6559 | 180146 |
| U1icnt | 2698 | 694 | 694 | 2698 | 5 | 20 |
| U2icnt | 1864 | 290 | 1842 | 1863 | 6 | 7 |
| U3icnt | 2108 | 2108 | 434 | 2108 | 5 | 6 |
| U1uvec | | | | | | |
| neutral | 0.16635 | 0.16651 | 0.16651 | 0.16635 | 0.13140 | 0.15147 |
| happy | 0.09731 | 0.09717 | 0.09717 | 0.09731 | 0.07738 | 0.09371 |
| sad | 0.11809 | 0.11856 | 0.11856 | 0.11809 | 0.14019 | 0.11820 |
| hate | 0.16420 | 0.16368 | 0.16368 | 0.16420 | 0.17097 | 0.17993 |
| anger | 0.11518 | 0.11425 | 0.11425 | 0.11518 | 0.12458 | 0.13003 |
| disgust | 0.17250 | 0.17262 | 0.17262 | 0.17250 | 0.19243 | 0.17438 |
| surprise | 0.16637 | 0.16720 | 0.16720 | 0.16637 | 0.16304 | 0.15228 |
| U2uvec | | | | | | |
| neutral | 0.17283 | 0.17278 | 0.17312 | 0.17284 | 0.17307 | 0.14432 |
| happy | 0.09686 | 0.09763 | 0.09671 | 0.09688 | 0.07707 | 0.08709 |
| sad | 0.11605 | 0.11663 | 0.11601 | 0.11605 | 0.12110 | 0.12285 |
| hate | 0.16121 | 0.16128 | 0.16130 | 0.16113 | 0.15901 | 0.18886 |
| anger | 0.11228 | 0.11215 | 0.11237 | 0.11227 | 0.12302 | 0.12316 |
| disgust | 0.17099 | 0.17031 | 0.17084 | 0.17099 | 0.19971 | 0.18004 |
| surprise | 0.16979 | 0.16922 | 0.16966 | 0.16984 | 0.14702 | 0.15369 |
| U3uvec | | | | | | |
| neutral | 0.16886 | 0.16886 | 0.16933 | 0.16886 | 0.13541 | 0.15547 |
| happy | 0.09975 | 0.09975 | 0.09930 | 0.09975 | 0.09764 | 0.08779 |
| sad | 0.11872 | 0.11872 | 0.11947 | 0.11872 | 0.15502 | 0.12221 |
| hate | 0.16088 | 0.16088 | 0.16035 | 0.16088 | 0.18060 | 0.20552 |
| anger | 0.11261 | 0.11261 | 0.11282 | 0.11261 | 0.09618 | 0.11536 |
| disgust | 0.17192 | 0.17192 | 0.17220 | 0.17192 | 0.17441 | 0.17048 |
| surprise | 0.16726 | 0.16726 | 0.16653 | 0.16726 | 0.16074 | 0.14316 |
| U1AII | 1.0 | 0.99999 | 0.99999 | 1.0 | 0.99485 | 0.99961 |
| U2AII | 1,0 | 0.99999 | 0.99999 | 0.99999 | 0.99380 | 0.99935 |
| U3AII | 1,0 | 0.99999 | 0.99999 | 1,0 | 0.98890 | 0.99922 |

The PAC connections show three disjoint users' ids: 414, 448, and 474 in MovieLens's mlsm dataset are connected to three larger MovieLens' datasets and cross-domain to Amazon users in the Digital Music dataset and Toys and Games dataset. The table shows the Amazon user id in the numeric format resulting from the programming assignment for computational convenience. The actual Amazon user id is in alphanumeric. To map the user id in numeric back to the actual user id requires a mapping function. For example, in Table 7, the U1 user has id 414 in the mlsm dataset PAC connects to user id 10354 in the Amazon Digital Music dataset. The actual user id 10354 in the Amazon Digital Music dataset is "A3CBNR1SZJJJDE".

## 7.1. Making Recommendations using Products in Amazon to PAC Users from MovieLens

As shown in Table 7, each user also shows the consumed items count in the respective dataset. For example, U1 user id 414 in MovieLens mlsm dataset has watched 2698 movies; whereas, U1 PAC user id 10354 in the Amazon Digital Music dataset has reviewed five music, and U1 PAC user id 97437 in Amazon Toys and Games dataset has reviewed 20 items. The following table depicts a sample of Amazon user id 10354 and 97437 have reviewed. The attribute names in Table 8 are as follow:

- Rid: reviewer id

- ASIN: Amazon Standard Identification Number. Except for books, the ASIN is the same as the ISBN. Almost all products on Amazon has their ASIN. It is a unique code Amazon assigns to a product that carries in the inventory.

- unixTime: Timestamp in Unix Julian time.

- Overall: rating score with scale 1 (lowest) to 5 (highest).

- Summary: short sentiment statement of a product review.

Table 8. Sample of Product Reviews by Amazon Users.

| Rid | ASIN | unixTime | Overall | Summary |
|-----|------|----------|---------|---------|
| 10354 | 9714721180 | 1023408000 | 5 | Has it really been 18 years |
| 10354 | B001237HCI | 1078358400 | 1 | Why, Oh Why |
| 10354 | B001237HCI | 1078358400 | 1 | Why, Oh Why |
| 10354 | B001UEYM5E | 1194134400 | 2 | Politically correct metal |
| 10354 | B0057PSUZA | 1061769600 | 2 | Flat |
| 97437 | 6301935063 | 1389312000 | 5 | a fairy tale |
| 97437 | 6303605672 | 1408752000 | 4 | a philosophical assassin |
| 97437 | 6305538972 | 1401753600 | 5 | eat or be eaten |
| 97437 | B00005ASOS | 1408752000 | 5 | an eternal test for believers |
| 97437 | B00005JPA6 | 1401408000 | 1 | black humor |

Table 8 listed five review samples for rid 10354 in Digital Music and Rid 97437 in Toys and Games. One product with ASIN 971421180 rated with 5 in the Overall score by Rid 10354l while the other products rated very low. Digital Music product with ASIN 971421180 becomes the right candidate for making a recommendation to other users who have a similar taste of Rid 10354. In Toys and Games Rid 97437, three product reviews were rated with five Overall scores. The first in the 5-score list with ASIN 6301935063 becomes the candidate for recommendation to other users. Hence, both Amazon products with ASIN 971421180 in Digital Music and ASIN 6301935063 in Toys and Games will recommend MovieLens U1 user id 414.

# 8. FUTURE WORK

The PAC method computes how similar a pair of objects' emotion profiles is through the AII. The object that involved may be a user or item. No restriction applies to any dataset for PAC computing if all involved datasets must conform to adapting the same protocol in obtaining the object emotion profile for items and users in the dataset. This study adopts Ekman's six primary human emotions and applies TAC as illustrated in [1] to EDR object for its emotion profile, IVEC, using the object's subjective text metadata. By taking the average of all IVEC values that a user has consumed, it yields the user's emotion profile, UVEC. Once IVEC and UVEC are known, one can apply PAC to compute the AII pairwise for each user in the source dataset against the other user in a target dataset regardless the target dataset is in the same or different information domains.

A future work item is to leverage the learned experience in the methodologies developed to make top-N recommendations in Emotion Aware Recommender (EAR) to extend its ability to Group Emotion Aware Recommender (GEAR). The following issues are in the future work plan: affective awareness in the grouping, group forming, group dynamic, and group decision making.

# 9. CONCLUSION

In this study, the concept of Affective Index Indicator (AII) of Affect Aware Pseudo Association Method (AAPAM) has further developed to make Pseudo Associate Connect (PAC) two or more disjoint users and items in different datasets across different information domains. Often, subjective writing such as product descriptions and sentiment reviews of products found in a product database are sources for emotion classification for product emotion profiles. Similarly, when users interact with products, it will leave a trail of interaction history embedded with users' preferences and choices to become the source for user emotion profile formulation. This paper demonstrated cross-domain recommendations making is possible using the AAPAM to PAC disjoint users across different domains. This paper showed how to join two disparate domain datasets for making recommendations seamlessly through detailed illustration. One can extend the PAC disjoint users from MovieLens to, for example, an advertisement database, then ads with similar emotion profiles can be recommended to MovieLens users.

# REFERENCES

[1]  Leung, John Kalung, Griva, Igor, & Kennedy, William G., (2020a) "An Affective Aware Pseudo Association Method to Connect Disjoint Users Across Multiple Datasets – an enhanced validation method for Text-based Emotion Aware Recommender", In: International Journal on Natural Language Computing (IJNLC) Vol, 9 (4), pp11-31.

[2]  Riedl, J., & Konstan, J., (2015) "MovieLens Latest Datasets (2015)", https://grouplens.org/datasets/movielens/latest/ [Stand: 2015; Zugriff: 19.04.2018].

[3]  grouplens.org, (2019) "Movielens ml-20m dataset readme", http://files.grouplens.org/datasets/movielens/ml-20m-README.html [Stand: 2019; Zugriff: 08.10.2019].

[4]  Bigdeli, Elnaz, & Bahmani, Zeinab, (2008) "Comparing accuracy of cosine-based similarity and correlation-based similarity algorithms in tourism recommender systems", In: Management of

Innovation and Technology, 2008. ICMIT 2008. 4th IEEE International Conference on. IEEE, pp469–474.

[5]    TMDb, (2018) "TMDb About", https://www.themoviedb.org/about?language=en/ [Stand: 2018; Zugriff: 11.05.2018].

[6]    Leung, John Kalung, Griva, Igor, & Kennedy, William G., (2020b) "Making Use of Affective Features from Media Content Metadata for Better Movie Recommendation Making", In: arXiv preprint arXiv:2007.00636.

[7]    IMDb, (2020) "Top 100 Greatest Movies of All Time (The Ultimate List)", http://www.imdb.com/list/ls055592025/ [Stand: 2020; Zugriff: 17.05.2020].

[8]    He, Ruining, & McAuley, Julian, (2016) "Ups and downs: Modeling the visual evolution of fashion trends with one-class collaborative filtering", In: proceedings of the 25th international conference on world wide web. International World Wide Web Conferences Steering Committee, pp507–517.

[9]    McAuley, Julian, & Ni, Jianmo, (2018) "Amazon Review Data (2018)", https://nijianmo.github.io/amazon/index.html [Stand: 2018; Zugriff: 11.05.2018].

[10]   McAuley, Julian, Targett, Christopher, Shi, Qinfeng, & Van Den Hengel, Anton, (2015): "Image-based recommendations on styles and substitutes", In: Proceedings of the 38th International ACM SIGIR Conference on Research and Development in Information Retrieval. ACM, pp43–52.

[11]   Pezoa, Felipe, Reutter, Juan L., Suarez, Fernando, Ugarte, Martín, & Vrgoč, Domagoj, (2016) "Foundations of JSON schema", In: Proceedings of the 25th International Conference on World Wide Web, pp263–273.

[12]   Sarwar, Badrul, Karypis, George, Konstan, Joseph, & Riedl, John, (2001) "Item-based collaborative filtering recommendation algorithms", In: Proceedings of the 10th international conference on World Wide Web. ACM, pp285–295.

[13]   Linden, Greg, Smith, Brent, & York, Jeremy, (2003) "Amazon. com recommendations: Item-to-item collaborative filtering", In: IEEE Internet computing, 7 (1), pp76–80.

[14]   Leung, John Kalung, Griva, Igor, & Kennedy, William G., (2020c) "Text-based Emotion Aware Recommender", In: arXiv preprint arXiv:2007.01455.

[15]   LeDoux, Joe, Phelps, Liz, & Alberini, Cristina, (2016) "What we talk about when we talk about emotions", In: Cell, 167, pp1443–1445.

[16]   Ho, Ilusca LL, Ai Thanh, Menezes, & Tagmouti, Yousra, (2006) "E-mrs: Emotion-based movie recommender system", In: Proceedings of IADIS e-Commerce Conference. USA: University of Washington Both-ell, pp1–8.

[17]   Ekman, Paul (1993): "Facial expression and emotion. In: American psychologist, 48 (4), pp384.

[18]   LeDoux, Joseph E., & LeDoux, Joseph E., (1995) "Emotion: Clues from the brain", In: Annual review of psychology, 46 (1), pp209–235.

[19]   LeDoux, Joseph E., (2012) "Evolution of human emotion: a view through fear", In: Progress in brain research. Elsevier, pp431–442.

[20]   Anderson, David J., & Adolphs, Ralph, (2014) "A framework for studying emotions across species", In: Cell, 157 (1), pp187–200.

[21]   Gu, Simeng, Wang, Fushun, Patel, Nitesh P., Bourgeois, James A., & Huang, Jason H., (2019) "A model for basic emotions using Observations of behavior in Drosophila", In: Frontiers in psychology, 10, pp7801.

[22]   Tayib, Saifulazmi, & Jamaludin, Zulikha, (2016) "An Algorithm to Define Emotions Based on Facial Gestures as Automated Input in Survey Instrument", In: Advanced Science Letters, 22 (10), pp2889–2893.

[23]   Plutchik, Robert, (2001) "The nature of emotions: Human emotions have deep evolutionary roots, a fact that may explain their complexity and provide tools for clinical practice", In: American scientist, 89 (4), pp344–350.

[24]   Fernández-Tobías, Ignacio, Cantador, Iván, Kaminskas, Marius, & Ricci, Francesco, (2012) "Cross-domain recommender systems: A survey of the state of the art", In: Spanish Conference on Information Retrieval.

[25]   Zheng, Yong, Mobasher, Bamshad, & Burke, Robin D., (2013) "The Role of Emotions in Context-aware Recommendation", In: Decisions@ RecSys, 2013, pp21–28.

[26]   Matsumoto, Kazuyuki, Yoshida, Minoru, & Kita, Kenji, (2018) "Classification of Emoji Categories from Tweet Based on Deep Neural Networks", In: Proceedings of the 2nd International Conference on Natural Language Processing and Information Retrieval, pp17–25.

[27] Sosa, Pedro M. (2017): "Twitter Sentiment Analysis using Combined LSTM-CNN Models", In: Eprint Arxiv.

[28] Pennington, Jeffrey, Socher, Richard, & Manning, Christopher D., (2014) "Glove: Global vectors for word representation", In: Proceedings of the 2014 conference on empirical methods in natural language processing (EMNLP), pp1532–1543.

## AUTHORS

**John K. Leung** is a Ph.D. candidate in Computational and Data Sciences Department, Computational Sciences and Informatics at George Mason University in Fairfax, Virginia. He has over twenty years of working experience in information technology research and development capacity. Formerly, he worked in the T. J. Watson Research Center at IBM Corp. in Hawthorne, New York. John has spent more than a decade working in Greater China, leading technology incubation, transfer, and new business development.

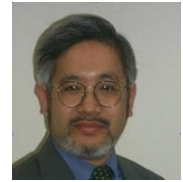

**Igor Griva** is an Associate Professor in the Department of Mathematical Sciences at George Mason University. His research focuses on the theory and methods of nonlinear optimization and their application to problems in science and engineering.

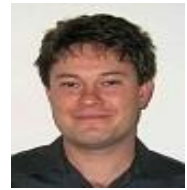

**William G. Kennedy**, PhD, Captain, USN (Ret.) is an Associate Professor in the Department of Computational and Data Sciences and is a Co-Director of the Center for Social Complexity at George Mason University in Fairfax, Virginia. He has over 10-years' experience in leading research projects in computational social science with characterizing the reaction of the population of a mega-city to a nuclear WMD event being his most recent project. His teaching, research, and publication activities are in modeling cognition and behavior from individuals to societies.

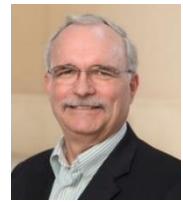